\documentclass{article} 
\usepackage{nips13submit_e,times}
\usepackage{url}
\usepackage{graphicx,color}

\title{Towards a  New Science of  a Clinical Data Intelligence}

\author{
Volker Tresp\thanks{
Corresponding author: Siemens AG, Otto-Hahn-Ring 6, 81739 Muenchen, Germany, volker.tresp@siemens.com.
 Published in: NIPS 2013 Workshop: \emph{Machine Learning for Clinical Data Analysis and Healthcare}, 2013.
}, Sonja Zillner, Mar\'{\i}a J. Costa, Yi Huang \\
Siemens AG, Corporate Technology\\
\texttt{firstname.lastname@siemens.com} \\
\And
Alexander Cavallaro, Peter A. Fasching,   Andr\'{e} Reis \\
University Hospital Erlangen\\
\texttt{firstname.lastname@uk-erlangen.de} \\
\AND
 Martin Sedlmayr, Thomas Ganslandt   \\
Medical Informatics, Friedrich-Alexander-University Erlangen-Nuremberg\\
\texttt{firstname.lastname@imi.med.uni-erlangen.de} \\
\AND
 Klemens Budde, Carl Hinrichs, Danilo Schmidt \\
 Charit\'{e} - Universit\"atsmedizin Berlin \\
 \texttt{firstname.lastname@charite.de} \\
\AND
Philipp Daumke \\
Averbis  GmbH \\
\texttt{philipp.daumke@averbis.com}
\And
Daniel Sonntag \\
DFKI GmbH \\
\texttt{Daniel.Sonntag@dfki.de}
\And
 Thomas Wittenberg \\
Fraunhofer Institute for Integrated Circuits\\
\texttt{thomas.wittenberg@iis.fraunhofer.de} \\
\And
Patricia G. Oppelt\\
Institut f\"ur Frauengesundheit GmbH \\
\texttt{Patricia.Oppelt@uk-erlangen.de} \\
\And
Denis Krompass \\
Ludwig Maximilian University of Munich \\
\texttt{dekromp@googlemail.com} \\
}

\nipsfinalcopy 

\begin{document}

\maketitle

\begin{abstract}


In this paper we define Clinical Data Intelligence as the analysis of  data generated in  the clinical routine with
the goal  of  improving  patient care.
  We define a
\emph{science} of a  Clinical Data Intelligence as a data analysis that permits the derivation of scientific, i.e.,  generalizable and reliable
results.
  We argue that a science of a  Clinical Data
Intelligence  is  sensible in the context of a  Big Data analysis, i.e., with data from many patients and  with
complete patient information.
 We discuss that
Clinical Data Intelligence requires the joint efforts of  knowledge engineering,  information extraction (from textual and other unstructured
data), and statistics and statistical machine learning.
 We describe some of our main results as conjectures and relate them to a recently funded research project involving two major German university hospitals.

\end{abstract}

\section{Introduction}
\label{sec:introduction}

 The healthcare sector has been identified as one of the main areas to  benefit from the recent trend towards Big Data
or large scale data analytics~\cite{McKinsey2013},  with an  increase in efficiency and improved patient care as desired outcomes. In this paper we assume the view point that clinical Big Data analytics needs to focus on the clinical decision
 processes.%
 We define Clinical Data Intelligence  as the analysis of data generated in the clinical routine with the goal  of
 improving  patient care.
  We define a
\emph{science} of a  Clinical Data Intelligence as a data analysis that permits the derivation of scientific, i.e.,  generalizable and reliable
results.
 We argue that a science of a  Clinical Data Intelligence  is particularly sensible in the context
of a  Big Data analysis, i.e., with data from many patients and  with complete patient information:
 Many patients are needed to capture the whole complexity of the medical domain and  complete information about
 each individual is needed to minimize the effect of confounders.
  Another point we want to make is that
Clinical Data Intelligence requires the joint efforts of  knowledge engineering,  information extraction from textual
and other unstructured
data, and statistics and statistical machine learning.
First, truthfully representing patient information  requires  ontologies and terminologies which have been developed in
recent years in medical knowledge engineering and in the context of the development of medical guidelines. Second, the
information on which the physicians base their decisions is often not contained in structural form but at
best in various textual reports. Information extraction is required to make this information available for a subsequent  analysis.
Due to the difficulties in extracting information from texts, such as medical reports, other source data, in particular images and increasingly OMICS data, can provide additional important information that can be made available by automated information extraction.
 Finally, statistics and statistical machine
learning provide a well established framework for the modeling and analysis of actions in a medical context. We argue
that Clinical Data Intelligence is the perfect field, where knowledge engineering, information extraction and
statistics and statistical machine learning can  benefit from one another. Potential applications are, first,
the prediction of actions (e.g.,  diagnoses or procedures)  to support a physician's decisions by modeling medical
practice, second, an analysis of  the benefits  of medical actions in terms of a final outcome and, third,  a system that provides the physician with indications, which of the potential actions under
consideration would generate the greatest patient benefit, realizing a true personalized medicine.
We discuss some of the  issues in the context  of a newly funded German research  project involving two major German
university hospitals.

The paper is organized as follows. In the next section we discuss medical decision support from a technical perspective and define technical requirements on a science of a Clinical Data Intelligence.
 In Section~\ref{sec:clinical} we discuss the state of medical processes and
documentation and discuss issues of information acquisition, semantic data models  and semantic annotations.  In Section~\ref{sec:ml} we discuss  statistical modelling and statistical decision analysis.  In
Section~\ref{sec:UseCases} we present  use cases of our project.
Section~\ref{sec:concl} contains our conclusions and an outlook.

\section{Requirements from a Technical Viewpoint} 
\label{sec:tech}

\subsection{Modeling Medical Decisions}

A medical decision support system would attempt to predict future  actions (e.g., decisions, procedures, treatments)
based on the available patient information.
To some degree,  the clinical state of the art in medical decision making  is represented in medical guidelines. These
reflect common best practice, are based on the results of well understood clinical trials and are at the core of the
implementation of an \emph{evidence based} medicine.
 There has been some effort to  formalize medical guidelines and make them computer accessible.
 Computerized clinical guidelines such as GEM, GLIF or Aredn Syntax have contributed to the formalization and automation of clinical data and knowledge.
Given the likelihood that physicians consider  even more information than what is  covered in the guidelines, we get as a
first conjecture:

\textbf{Conjecture~A:} \emph{A Science of a Clinical Data Intelligence requires that the models and algorithms can use
the same information  which the physician is using at the point of the decision. Medical  guidelines provide  insight
into what that information should be.}

In terms of Big Data the conjecture implies that we need \emph{variety}, i.e., a representation of the complex context
of a patient.  To be able to represent this information in sufficient detail reflecting the complexity of medicine we
get the next conjecture:

\textbf{Conjecture~B:} \emph{In Clinical Data Intelligence, the models have to be able to truthfully represent medical
facts i.e., work with the medical language as represented in medical ontologies and terminologies.}

In other words, the algorithms need to understand the expressiveness of medical information, e.g., as required for
medical guidelines. Here one can build on many years of previous work on medical knowledge representations, i.e., the
large corpora of medical ontologies, taxonomies and dictionaries, e.g., SNOMED (Systematized Nomenclature of Medicine), ICD (International Classification of Diseases), RadLex (covering radiology-specific terms), and FMA (Foundational Model of Anatomy ontology).

\subsection{Improving Decisions}

Outcome information can be used to guide  physicians towards more informed decisions. The basis for such an analysis would
be a model
for the probability of an outcome, given patient state ---as known by the physician---  and given the action
performed.
To identify such a model,
we need rich data reflecting a variety of patient information and a variety of actions.

\textbf{Conjecture~C:} \emph{A Science of a Clinical Data Intelligence requires that the collected patient database is
sufficiently rich and reflects a variety of patients and decisions.}

Naturally, if two medications are always prescribed jointly, then there is no way to learn the effect of one without the
other; correlated inputs make the identification of a model difficult and we need rich data to break the correlations.
In terms of Big Data this means that in addition to variety,  we need \emph{volume}!
Combining Conjectures A and C leads to:

\textbf{Conjecture~D} \emph{A Science of a Clinical Data Intelligence requires  a Big Data analysis.}

 Confounders are variables
that are taken into account in the decision but are not reflected
in the data and are the main reason  why an outcome-oriented  analysis of actions is difficult. 
From  a modelling point of view  confounders are latent variables correlated both with action and outcome.  The
most compelling example is if a placebo medication is only given to currently healthy people  but the health
states of the patients  is not documented. An analysis would then see  a strong  correlation between a positive
outcome and medication.
 The solution  is to \emph{control the confounders}:
  A Big Data approach  would be to include all variables in the data that are candidates for being 
 confounders, in the example one should include at least  the   health status of the patient at the time of the administration of the placebo medication. 
The conclusion is that to minimize the effect of confounders, Conjecture~A needs to be
fulfilled.\footnote{Unfortunately the process is not fool proof; including only a subset of confounders might in rare
cases increase bias~\cite{Pearl2009}}

The field of epidemiology has developed sophisticated  methods to analyse causal effects in observed data, e.g.,
retrospective cohort studies,  and propensity score matching.
In particular a generalization of the latter could be of interest since it is well suited to the high-dimensional data
situation in Clinical Data Intelligence. Unfortunately, not all confounders can be made visible. The correction for
hidden confounders is an active area of research~\cite{DBLP:conf/uai/ListgartenKSH11}.

Despite the difficulties in working with observed data from clinical care there are also advantages. First, one gets
the data for free (not quite) and the data are collected in a realistic clinical setting and not in a somewhat
artificial study situation. Also other sources for introducing bias, such as
selection bias and  information bias,   might  not be as much of a problem.

\section{Clinical Data}
\label{sec:clinical}

\subsection{A Clinical Process}

A patient enters the hospital with one or several complaints and a  
goal  is to understand the condition and the problems of a patient via anamnesis and diagnostics steps, possibly  including radiological imaging, blood counts,  histological examinations, and other tests. A clinical stay typically also include interventions to stabilize and  improve the patients status,  e.g.,  via surgery and the administration of medications.  Analysis and interventions are not strictly sequential steps but repeated iteratively:  the effect of an intervention might be checked by some form diagnostics which then might lead to changes in treatments. Decisions are made by physicians individually as well as in a larger group of medical experts. At discharge the patient is informed about further required treatments,  and  recommended life style changes are being discussed.

\subsection{The State of Medical Documentation}

Ideally, the medical processes should be documented such that each step may be reconstructed at a later time. Current practice is still far from realizing this goal and the main purposes of a documentation are legal obligations and billing. The clinical relevance of billing information is somewhat limited since the coding is often performed by personnel not directly involved in the case and it is done near the end of or even after the hospital stay.

 Unfortunately,  only a small part of the information is available in structured form, as for example lab results (e.g., blood counts) or  diagnoses and procedures.
  It would be highly desirable that reports and records would be provided in coded format, as parts of electronic health records (EHRs).
 The situation is improving,   but structured documentation by using codes is only slowly being adopted in clinical practice.
 

 It has been estimated that up to 80\% of patient information is only available in unstructured textual form and that this information is perceived to be more trustworthy and complete, when compared to coded information.
Thus it seems unavoidable to analyse the textual reports to gain the information required for Clinical Data Intelligence. We formulate this as a conjecture:

\textbf{Conjecture~E:} \emph{A significant portion of the patient information is only available in a textual format. Thus information extraction from text is an essential part of Clinical Data Intelligence.}

Information is also available as   source data or raw data, in particular as   images generated in  radiology and pathology. In recent years there has been quite some effort to automatically annotate images (for example,~\cite{Seifert2009}). Although it might be debatable if these algorithms have reached the level of an experienced expert (likely not), from a Clinical Data Intelligence point of view one might argue that the certainty and the quality of the information that can be extracted automatically from the source  data is higher than the quality and certainty of the information that can automatically be extracted from the textual reports. For the foreseeable future, one should exploit both, the
source data and the textual reports, jointly,  such as in \cite{Dankerl2013}.

\subsection{Semantic Annotation of Texts and Images}

Information extraction from text is typically performed by a combination of natural language processing  and  machine learning. Current state of the art methods are able to reliably detect key entities in texts;   the extraction of statements from texts, much more valuable for describing the patient status,  unfortunately is less reliable.
 A feasible solution is that the degree of uncertainty  is represented in the annotations and the consecutive processing steps take this uncertainty into account. The analysis of medical texts is particularly challenging since sentences are often incomplete, use  clinic-specific terms and contain an abundance of  negations.

The automatic analysis of the source data (radiology, pathology, EKG, EEG, ...) has made great progress in recent years. For example as part of THESEUS MEDICO\footnote{http://www.theseus-programm.de/}, it was possible to locate and measure major organs as well as healthy and diseased lymph nodes~\cite{DBLP:conf/kes/MollerESSGCD10}.  We expect that  our  project will greatly benefit from  the inclusion of   imaging features and semantic image annotations.

\subsection{Speaking the Right Language: a Semantic Data Model}

Considering the complexity of the semantic annotations of unstructured data and also the complexity of the patient information required for the medical guidelines, a simple relational schema might be insufficient to represent the data. In our project we develop an ontology-based semantic triple store to represent complex patient information and to make this information accessible. The ontologies in these triple stores are derived from standard medical ontologies and taxonomies such as  SNOMED (Systematized Nomenclature of Medicine), ICD (International Classification of Diseases), RadLex (covering radiology-specific terms), and FMA (Foundational Model of Anatomy ontology). Note, that a triple store  does not only provide schema information but is also  used  to represent relevant medical background knowledge.

Care must be taken such that a sensible compromise is formed between ontological expressiveness and complexity on the one side and scalability and usability on the other side.

%

\subsection{A Research Database}

 In our project we work with a research database  that stores the available patient data and makes it available for research and  model development. The database contains all  structured patient data,  all textual documents, all annotations of texts and of other unstructured data, as well as   pointers to the sources.  The research database is complemented by the triple store described in the last subsection for representing more complex semantic annotations. 
 
 Typically, patient data is scattered over many systems in a hospital such as radiology information system, laboratory information system and the electronic patient record. This is why considerable effort is required to develop connectors to these data storage facilities to import the data into the research database and to map this heterogenous data onto a common terminology.

The research database  in our project is based on i2b2, which is has been developed by Harvard and MIT and has been used in a number of translational biomedical research projects.\footnote{https://www.i2b2.org/software/index.html}   i2b2 provides a number of operational and quite useful plugins.

Only in passing we want to mention the issue of privacy, which often requires anonymization. Researchers should only have access to anonymized  data and corresponding access control needs to be implemented. Whereas the anonymization of structured data is relatively simple, the anonymization of unstructured data requires considerable effort and quality control in our project. The goal is a  HIPPA conform information exchange~\cite{McKinsey2013}.

\subsection{The Increasingly Important Role of OMICS Data (Molecular Data)}

Personalized medicine is typically associated with the increasing clinical importance of OMICS data (genomics, proteomics, metabolomics, ...). Genetic testing is already performed as a molecular test in combination with the administration of certain drugs, as companion diagnostics. The susceptibilities to certain types of breast cancer are determined by the analysis of genetic mutations   and cancer states are determined by expression profiles of cancerous tissue. Next generation sequencing will make genetic testing inexpensive and interesting to wider classes of disease treatments, in particular in connection with various types of cancer.
OMICS data is expected to play an increasingly  important  role in the  clinical medicine of the future, potentially producing terabytes of data for each individual patient~\cite{Stanford2012}. The analysis of the dependencies between genetic factors (genotype),  expression data  and clinical phenotypes will be investigated in our project.

\subsection{Outcome Information}

Clinical outcome is not always well documented and new policies attempt to improve the situation.
Readmission within a certain period of time (typically a month) is sometimes taken for a
negative outcome.
Alternatively one might define a hospital stay of more than a certain number of days as a negative outcome, where the threshold is disease-related-group (DRG) specific.
In some cases, for example after a kidney transplantation or mastectomy, the patient is closely observed, potentially  over lifetime, and outcome is monitored.

%

\section{Modelling Decisions with Machine Learning}
\label{sec:ml}

The common way to model patient data is to build a patient model, typically in form of a probabilistic (causal)  Bayesian network, then enter the available evidence (patient properties, symptoms), and perform probabilistic inference to infer about the diagnoses and procedures.  We consider this a model-based  approach. Naturally these models can become increasingly complex and  require explicit expert modelling. With increasing model size, inference can become very expensive and often approximate forms of reasoning are applied. Another  potential problem with this approach is that experts need to agree on quantified dependencies (causal structure, conditional probabilities) and that recommended actions might not be in accordance with general practice. A serious and  successful effort in Europe is the PROMEDAS decision support system~\cite{DBLP:conf/aime/WemmenhoveMWLKN07}.

Clinical guidelines and also clinical processes in general work differently. They consider the state of the patient,  and then prescribe what actions to take (do additional tests, do a procedure, prescribe a particular medication). A lot of analysis and thinking has gone into the development of the guidelines, but, at the end, they are conditional instructions and work in a direct manner.
Clinical studies also pursue a direct approach: Certain patient properties are recorded and the effect of a treatment is evaluated.

We believe that direct approaches are better suited than model-based approaches  for Clinical Data Intelligence where the basic assumption is that data speaks for itself,  leading to the possibility of gaining novel insights.

An important goal in our project is the prediction of an action. Note,  that this can be quite useful in a clinical setting and that this task is not affected as much by confounders: essentially the more relevant information is available, the better the quality and the explainability of the  prediction.

There is a certain analogy to recommendation engines where the state of the patient  corresponds to  the state of the customer of a shopping platform  (items which were  liked and disliked in the past by the customer) and where the predicted actions correspond to the recommendations to be made (advertising shown to the customer).

In our project we  apply matrix and tensor factorization  models,  which fit well to the relational structure of
the medical data situation with high-dimensional sparse data~\cite{Tresp:09,Nickel2012} and have shown state-of-the art performances on  ontologies with several million  entities,  where the likelihood of  $10^{14}$ possible relations
could be predicted. Translated to  Clinical Data Intelligence: the likelihood of all possible actions and diagnoses for a patient can be obtained in one highly efficient computational step.

\begin{figure}[htb]
\centering
\includegraphics[scale=0.85]{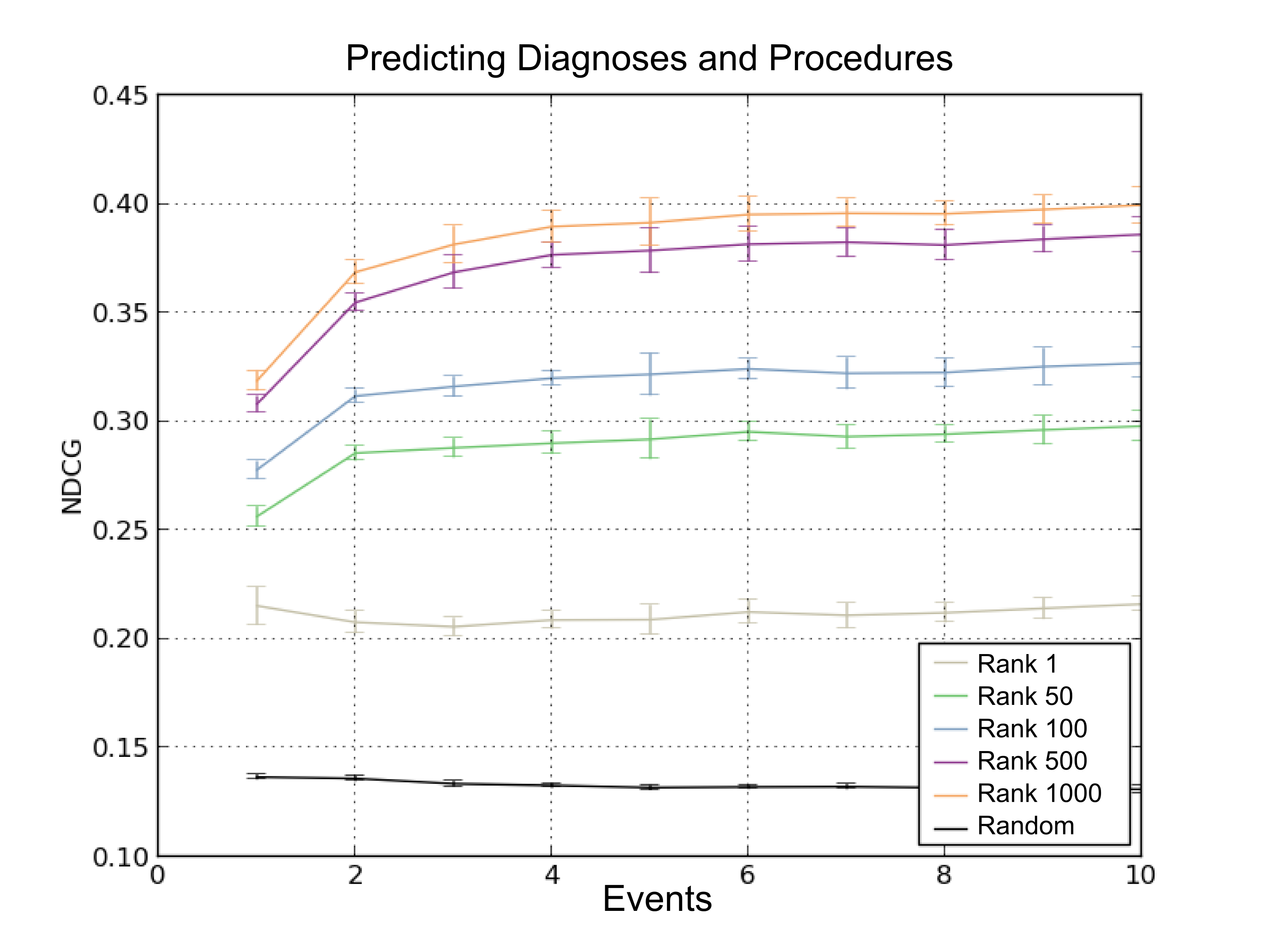}
\caption[adsd]{Data from 10000 patients were used. We considered
2331 possible diagnoses,
1634 possible procedures,
2721 possible lab results,
209 possible therapies and
281 general patient data.  In total the data contained  $5.9$ million facts. We predicted the next decision (diagnosis, procedure) as a function of the information
available for each patient.  Plotted is the NDCG score (a popular score for evaluating ranking results~\cite{DBLP:conf/sigir/JarvelinK00}) as a function of the information available for each patient (a
large number is  desirable).  An event corresponds to an instance in time where patient data is recorded.
With increasing information, the prediction improves. We see plots for different
approximation ranks: the highest rank gives best scores which reflects the high degree of data complexity.
}
\label{fig:rescal2}
\end{figure}

Figure~\ref{fig:rescal2}  shows first experimental results. We predicted procedures and diagnoses based on  patient
information   and tested if our predictions coincide with actually given diagnoses and procedures.
Actually given diagnoses and procedures on test data  were statistically in the top three of our predicted diagnoses and procedures (we used data from 10000 patients and considered 3965 procedures and diagnoses).

\section{Use Cases}
\label{sec:UseCases}

We now describe some use cases of our project and describe  specific goals.

\subsection{All Data from All Patients}

The first use case is unspecific in the sense  that we want to cover all patients in a clinic, in disregard of their specific problems. In the research database we collect various  data from all clinical departments; this includes core patient data, diagnoses, procedures, medications,  transfusions, lab values,  lab reports, tumor board decisions, DRG codes, surgery data (surgery core data, diagnoses, procedures, material), OMICS data, pathology reports, radiology reports, and release reports. In addition we include semantic annotations and pointers to the source data (radiological and pathological images, time series data).  We use this data to develop an integrated data model for representing patients.

With a perspective of a personalized medicine,  we model the patient in her/his full complexity and accordingly can provide patient specific recommendations.  We cross department boundaries and  obtain a global model  of the clinical data and of the clinical decision processes.

\subsection{Breast Carcinoma: Data from Different Modalities}

Breast carcinoma is one of the most common malignancies in women. 
Relevant events are screening, diagnosis, therapy and follow-up care.   
The classical screening procedure is a low-energy breast X-ray (mammography), sometimes supplemented with ultrasound and MR imaging. Genetic counseling might suggest a test for certain gene  mutations  for high risk patients. Mutations in  the breast cancer susceptibility gene 1 (BRCA1) and breast cancer susceptibility gene 2 (BRCA2) predispose to breast cancer (85\% lifetime risk) and ovarian cancer (27-44\% lifetime risk). Hereditary breast cancer is associated with a poorer survival, and hereditary ovarian cancer with a better survival than their sporadic counterparts, an observation which can affect the prevention, surveillance and treatment strategies. Carriers of such mutations are at high risk of developing other forms of cancer such as  pancreas cancer. The early detection of the mutations may affect strategies for clinical management of these cancers.  In high risk cases the patient might decide to take preventive medications (chemoprevention)  or even decide for   a prophylactic surgical removal of the breasts. A diagnosis is performed based on radiological imaging, histological examinations, and gene expression profiles of the cancerous tissue, including  the Oncotype DXTM Breast Cancer Assay which looks at the properties/expressiveness of  21 genes.  As part of a diagnosis the appearance of the cancer cells is evaluated (grade), the breast cancer stage is determined, and the receptor status might be determined. Receptors (estrogen receptors (ER), progesterone receptor (PR), and HER2 receptors) on the surface of cancer cells  are binding sites for chemical messengers (hormones) and determine the optimal therapy. Surgery is performed to  physically remove the tumor, typically along with some of the surrounding tissue.   Adjuvant therapies include radiation therapy and  chemotherapy. Patients with estrogen receptor positive tumors will typically receive hormonal therapy after chemotherapy is completed.

Data driven decision support can support the different disease stages.    A specific use case relates to the issue that in Germany and other countries, genetic testing is only reimbursed when the probability of existence of a genetic mutation associated with an increased risk of breast cancer is above a certain threshold. Therefore, predictors not only of the likelihood of development of breast cancer, but also of the probability of having such mutations become of utmost importance, and is one of the goals of the project. There is strong evidence in the literature supporting the existence of imaging differences between mammograms of patients with and without such genetic background.
However, and in spite of the advances in modern image analysis techniques, these differences have not been exploited until now. It is therefore one of our main goals in this use case to study these differences in depth.

As another use case, documented evidence supports the claim that imaging differences exist between inherited genetic breast tumors  and sporadic breast tumors. These differences can often be subtle image features that might not be discernable by humans, but might still be distinguished through automatically computed image descriptors. We will leverage our experience with imaging features together with additional information such as family history and histopathology results to provide the best possible prediction of the existence of a mutation-related tumor, and to provide robust and reliable decision support in the framework of breast cancer.

Overall the dependencies between image analysis, genetic features and gene expression data, and general patient data will be a focus of our project.   We will work with  data collected in clinical routine and data  from clinical studies  (e.g., the Bavarian breast cancer study). Data from clinical studies are collected from a number of hospitals permitting comparative analysis. The data contains 1,000,000 genetic factors plus 44,000 expression data sets obtained from tumor tissue. Radiological images are described by several hundred features extracted from the images.  Of special interest here is the determination of risk factors, the evaluation of the therapy and the prediction of side effects.

The imaging work will build upon our own previous  studies on tumor detection and characterization, where we exploited a number of tissue descriptors of different kind, which can directly be applied to describe lesions in the breast. Low--level automatic descriptors, which proved to be informative in our previous work, include the fast 3D Haar-like features, statistical moments of intensity values and intensity histograms, image moments of Hu and Zernike, objectness features that track tissue blobness, vesselness and planeness, the convergence index features, sample symmetry tracking features, and various shape descriptors of the candidate's surface~\cite{DankerlLiver2013}. In addition to automatic descriptors, human input with sample characteristics which are not easily computable from the data is often valuable, too.

\subsection{Nephrology with Temporal Data and Outcome Data}

Kidney diseases are causing a significant financial burden for the German health system. It is estimated that alone the treatment of end-stage renal disease (ESRD) with chronic renal replacement therapies accounts for more than 2.5 billion Euros annually, and the incidence of dialysis-dependent renal insufficiency is rising by 5-8\% each year.\footnote{http://www.ncbi.nlm.nih.gov/pubmed/11792760} Despite progress in diagnosis, prophylaxis and therapy of chronic kidney diseases, renal transplantation remains the therapy of choice for all patients with ESRD. Kidney transplantation leads to a significant improvement of quality of life, to substantial cost savings and most importantly to a significant survival benefit in relation to all other renal replacement therapies. 2850 kidney transplantations were performed in Germany in 2011 but more than 8000 patients are registered on the waiting list for a kidney transplant.\footnote{http://www.dso.de/organspende-und-transplantation/transplantation/nierentransplantation.html} With excellent short-term success rates, nowadays the reduction of complications and the increase of long-term graft survival are the main goals after transplantation, especially considering the dramatic organ shortage. It is not only important to reduce - or better avoid - severe and/or life-threatening complications such as acute rejection, malignancy and severe opportunistic infections, but it is also of utmost importance to ameliorate the many other serious side effects, which increase cardiovascular risk, decrease renal function, necessitate costly co-medication or hospitalizations and also have an impact on the quality of life after successful transplantation. Although renal transplantation is considerably cheaper than regular dialysis treatment it is a complex and costly procedure. Due to the outlined complexities, patients should remain in life-long specialized posttransplant care. Patients have not only to take immunosuppressants, but also have to take numerous drugs for prophylaxis and treatment of preexisting and/or concomitant diseases, which are at least in part aggravated by the immunosuppressants. As a consequence most patients have to take 5-10 different medications. The many drugs and the multiple side effects of the routinely administered medication cause a substantial cost burden. There is not only a medical need but also a financial necessity to reduce side effects, diagnostic procedures, therapeutic interventions, hospitalizations and ultimately improve patient safety. This will directly lead to a better quality of life, cost savings and better allocation of medical resources. Aim of this work is to systematically investigate drug-drug interaction (DDI) and adverse drug reactions (ADR) in patients after renal transplantation. In the first step we want to assess the frequency of such events in the well-structured database of our transplant centre. The descriptive analysis and harmonization provides the basis for further investigations. In a second step we are aiming to find new DDI and previously unknown ADR utilizing modern data-mining techniques. We will investigate their effect on key outcome parameters such as patient and graft survival, renal function as well as hospitalizations. Lastly, we will implement a clinical decision support system directly into the electronic patient file, in order to prevent dangerous DDI, reduce dosing errors and provide the physician and patient with timely and adequate information on new prescriptions. In summary, we aim to investigate ADR in a well maintained database in a thorough and systematic way to realize a clinical decision support system for the electronic patient record.

\section{Conclusions}
\label{sec:concl}

In the paper we have addressed Clinical Data Intelligence and postulated a number of conjectures we believe are
important. Maybe the most important one is that a sensible  analysis of data collected in a daily routine implies a Big Data analysis.  We described under which conditions we believe scientific conclusions can be drawn from the data. We
believe that much can be done already but to harvest the full power of a Clinical Data Intelligence, we need an
improved clinical data situation. Once the information can be unlocked  via a semantically rich, diverse and large data set, the potential benefit will be tremendous.

Even more data might be considered.
Data from social networks, forums, and blogs can provide information about  patient views and problems on many health issues.
Growing amounts of background data becomes available in digital formats, examples are Wikipedia, digital textbooks, and medical publications. These can
act as references  and  sources of  medical background information and as supplementary information supporting clinical decisions.
Other interesting sources of information are claims data since they might contain data over the lifetime of a patient permitting the analysis of outcome and cost effectiveness.

There is also the issue of collecting data across clinic borders. This is highly desirable but currently poses great technical and procedural problems.  Hospitals are, for the right reasons, very sensitive with their patient data. Still it is desirable that data sharing should be  the default and not the exception~\cite{McKinsey2013}. Registries are centralized patient data collections  for specific diseases. These can be quite voluminous but only contain very basic  information. If the data gets richer they will increasingly play a role in Clinical Data Intelligence.

Clinical Data Intelligence will certainly also require new business models where the clinics open up to an economy of service and tool providers. An era of open innovation is under way and   medical App platforms are being developed (i2b2\footnote{https://www.i2b2.org/software/index.html}, Mirth Results\footnote{http://www.mirthcorp.com/products/mirth-result}, OpenMRS\footnote{http://openmrs.org/}, WorldVistA EHR\footnote{http://worldvista.org/World\_VistA\_EHR/voe\_features}).

Finally, one needs to consider usability. How should information be presented in the medical workflow such that it optimally supports the decision processes and is accepted by the clinicians?   New smart devices might be increasingly used for data capture, but could also lead to novel point-of-care interfaces. This aspects needs special attention since solutions are most valuable if they are becoming part of  clinical practice.

\section*{Acknowledgement}

The project receives funding from the German Federal Ministry of Economics and
Technology; grant number 01MT14001.
 \bibliographystyle{unsrt}
{\small
 \bibliography{KDInips2013-arXiv}
}

\end{document}